\documentstyle[manuscript,aps]{revtex}
\tighten
\title{Chaos in the one-dimensional gravitational three-body problem}

\begin{document}

\author{Jarmo Hietarinta${}^{a,b}$ and Seppo Mikkola${}^{c,b}$}
\address{ ${}^a$Center for Nonlinear Studies, Los Alamos National Laboratory,
Los Alamos, NM~87545\\
${}^b$Department of Physics, University of Turku, 20500 Turku, Finland
\footnote{Permanent address} \\
${}^c$Turku University  Observatory, 21500 Piikki\"o, Finland}

\maketitle

\begin{abstract}
We have investigated the appearance of chaos in the 1-dimensional
Newtonian gravitational three-body system (three masses on a line with
$-1/r$ pairwise potential). In the center of mass coordinates this
system has two degrees of freedom and can be conveniently studied
using Poincar\'e sections.  We have concentrated in particular on how
the behavior changes when the relative masses of the three bodies
change.  We consider only the physically more interesting case of
negative total energy.  For two mass choices we have calculated 18000
full orbits (with initial states on a $100\times 180$ lattice on the
Poincar\'e section) and obtained dwell time distributions.  For 105
mass choices we have calculated Poincar\'e maps for $10\times 18$
starting points.  Our results show that the Poincar\'e section (and
hence the phase space) divides into three well defined regions with
orbits of different characteristics: 1)~There is a region of fast
scattering, with a minimum of pairwise collisions. This region
consists of `scallops' bordering the $E=0$ line, within a scallop the
orbits vary smoothly. The number of the scallops increases as the mass
of the central particle decreases.  2)~In the chaotic scattering
region the interaction times are longer, and both the interaction time
and the final state depend sensitively on the starting point on the
Poincar\'e section.  For both 1) and 2) the initial and final states
consists of a binary + single particle.  3)~The third region consists
of quasiperiodic orbits where the three masses are bound together
forever.  At the center of the quasiperiodic region there is a
periodic orbit discovered (numerically) by Schubart in 1956. The
stability of the Schubart orbit turns out to correlate strongly with
the global behavior.
\end{abstract}

\section{Introduction}
The three-body problem is one of the fundamental problems of dynamical
systems. It has a long tradition, even in connection with chaos
research: This is the problem that led Poincar\'e to make his famous
observations of chaos.

Most of the chaotic systems studied during the past twenty years have
been compact systems, where a good understanding of e.g. the way
chaos enters has been obtained. Recently more interest has been paid
to scattering problems. One popular problem is the
scattering of a particle on the plane from a fixed localized object
[1,2]. More closely related to the present topic is the work on
reactive scattering in chemistry [3].

The gravitational Newtonian three body scattering problem is
interesting and special due to the following three properties: It is
noncompact, the interaction is long range, and it is not obtainable as
a small perturbation of an integrable system. These properties mean
that many of the important theorems characterizing chaos are not
directly applicable to this system. [In the present gravitational case
the interaction is attractive, and our results cannot be translated to
the corresponding electrostatic case where some objects interact
repulsively.]

The general three dimensional three-body problem has been studied
extensively by numerical simulations [4] and by analytical studies
[5,6] (for a review see [7]).
However, in full generality this problem is still too complicated for
a systematic analysis.  This has led to the study of various
simplifications or restrictions of it; these include:
\begin{enumerate}
\item The restricted three-body problem: Three dimensional, but
one of the masses is assumed to be zero. The motion of the two massive
bodies is integrable, but the motion of the third body in the field of
the other two is not.
\item Three dimensional with finite nonzero masses but with
special symmetries. [E.g.  two masses symmetrically on the
$(x,y)$-plane with the third moving on the $z$-axis [8].]
\item The planar three-body problem: motion is restricted to a
plane, masses are free.
\item The rectilinear three-body problem: motion is restricted
to a line, masses are free.
\end{enumerate}
The last, simplest case, which is the topic of this article, is still
complicated enough to show a rich variety of phenomena. It also has a long
history [9-11] but it has still not received a definitive analysis.

Recently the present authors made a comprehensive study of the
rectilinear three-body system when the masses are equal [12,13]
(hereafter referred to as papers I and II).  We have also obtained
some results for arbitrary masses [14] (paper III). In the present
paper we review the previously obtained data and then go on to discuss
in detail the situation where the masses are arbitrary. We will mainly
consider the more interesting negative energy case, the dynamics with
nonnegative energies and equal masses was discussed in paper II.

There are various points of view that one can use when studying a
dynamical system. In the present case the system can be reduced to a
conservative system with two degrees-of-freedom and therefore we can
look at it through conventional method of Poincar\'e sections. Since
the system allows scattering we have two other quantities that can be
used to characterize the dynamics. First we can look at the dwell
time, i.e.  the time all particles stay together and interact
strongly before the triplet again breaks up. Secondly we can
characterize the scattering process by the types of initial and final
states. All of these methods are used here.

We start in Sec. 2 by presenting the system and showing how the
initial data/Poincar\'e section is defined so that it is possible to
include all orbit types.
In Sec. 3 we get a preview of the richness of the system by looking
at a set of typical trajectories.
In Sec. 4 we take the first global view by looking for two particular
mass choices how the dwell time depends on the initial values.
In the remaining Sections we consider the full mass freedom, first
in Sec. 5 through the stability of the periodic Schubart orbit.
In Sec. 6 we present a set of 105 Poincar\'e sections computed
for a lattice of mass values. The Poincar\'e sections show interesting
transitions as the masses change.
In Sec. 7 we discuss how the stability of the periodic Schubart orbit
influences the initial and final state types.

\section{ The system}
In this section we first describe the system and its reduction
to two degrees of freedom. Then the definition of the
Poincar\'e section (and initial values for numerical computations) is
discussed in detail.
Finally we describe how the Hamiltonian is regularized into a form
that is useful in numerical integrations and discuss the accuracy of
the computations.

\subsection{The Hamiltonian}
The mass configuration is given in Fig.\ \ref{fig1}.  We have three masses,
labeled $i=1,\, 0,\, 2$ from left to right, with masses $m_i$ at
positions $x_i$. The Hamiltonian is
\begin{equation}
H={\textstyle\frac12}\sum_{i=0}^2\frac{w_i^2}{m_i}
-\sum_{i<j}\frac{m_im_j}{|x_i-x_j|},
\end{equation}
where $w_i$ are the momenta canonical to $x_i$ and we have normalized
the system so that the gravitational constant $G=1$.

The attractive nature of the interaction leads to collisions, which
seem to be singular. For a pairwise collision the singularity is not
essential and there is a standard method by which the singularity can
be regularized [15,16]. Intuitively the singularity can be opened if
we consider colliding masses with a small but nonzero impact
parameter. When they are close they will not collide head on but
revolve rapidly around each other and emerge from the collision region
with momenta reversed in their center of mass frame. Thus when we let
the impact parameter approach zero we get an attractive potential with
a reflecting hard core. As a result we have a `reflection' in which
the bodies cannot change their order on the line.
The rectilinear two-body collision can also be
solved analytically when the distance to the third body is much larger
than that between the two colliding bodies. [In that case we get two
approximate two-body systems: one for the two colliding pair, and one
for the third body and the center of mass of the colliding pair.]
The triple collision, however, is a genuine
singularity and cannot be regularized for generic masses [17].

Since we have only pairwise interactions we can go to the
center-of-mass frame and eliminate one degree of freedom.
The canonical transformation from $(x_i,w_i)$ to $(q_i,p_i)$ is
defined as follows [$M=m_1+m_0+m_2$]:
\begin{equation}
q_1=x_0-x_1,\,q_2=x_2-x_0,\,
q_0=(m_1x_1+m_2x_2+m_0x_0),
\end{equation}
\begin{equation}
w_1=-p_1+{\textstyle\frac{m_1}M}p_0,\,w_2=p_2+{\textstyle\frac{m_2}M}p_0,\,
w_0=p_1-p_2+{\textstyle\frac{m_0}M}p_0,
\end{equation}
and after omitting the center of mass term the new Hamiltonian becomes
\begin{equation}
H={\textstyle\frac12}\left(\frac1{m_1}+\frac1{m_0}\right)p_1^2
-{{p_1p_2}\over {m_0}}
+{\textstyle\frac12}\left(\frac1{m_2}+\frac1{m_0}\right)p_2^2
-{{m_0m_1}\over q_1}-{{m_0m_2}\over q_2}-{{m_1m_2}\over {q_1+q_2}}.
\end{equation}

As shown before the particles preserve their order on the line and
therefore we always have $q_1\ge0$, $q_2\ge0$.  The motion takes place
in the first quadrant of the $(q_1,q_2)$-plane with attracting walls
at coordinate axes, see Fig.\ \ref{fig2}. There is also attraction
from the nonphysical wall at $q_1+q_2=0$, which can be reached only at
the origin which is an essential singularity.  In Fig.\ \ref{fig2} the
Poincar\'e section (defined in Sec. 2.2) is given by the line
$q_1=q_2$, i.e.  when the system crosses this line it crosses the
surface of section.  The motion illustrated in Fig.\ \ref{fig2} is a
typical fast scattering orbit.  The arrows show the direction of
motion of the point $(q_1,\;q_2)$ (since the system is time reversal
invariant the direction of the arrows can be reversed). The trajectory
is unbounded in both directions of time, which is the typical
situation. There are three crossings of the Poincar\'e section in this
example, two of which are very close to each other (on the surface of
section) but not equal.

Let us note that there is one special solution, the so called {\it
homographic} solution [18], for which the ratio
\begin{equation}
\tau=\frac{q_1}{q_2},
\end{equation}
is time independent. Clearly this leads to a triple collision.  To
find this solution and the particular ratio, $\tau=\tau_h$, one
substitutes $q_2=q_1/\tau_h$ and $p_2=\beta p_1$ into the equations of
motion obtained from (4) and requires that the two pairs of equations
for $q_1$ and $p_1$ are consistent.  [In fact both pairs lead to the
one-dimensional two-body problem $\ddot q=\tilde M q^{-2}$, which is
analytically solvable, and yields a periodic orbit. Thus this triple
collision orbit is regular, but any nearby orbit is completely
different.]  The consistency condition yields two equations for
$\tau_h$, $\beta$ and the masses $m_i$. After eliminating $\beta$ and
parametrizing
\begin{equation}
\tau_h=\frac{1+z}{1-z}.
\end{equation}
one can write the final condition in the nice form
\begin{equation}
{{-1\cdot m_1+z\cdot m_0+1\cdot m_2}\over{m_1+m_0+m_2}}
={{z^5-2z^3+17z}\over{z^4-10z^2-7}}.
\end{equation}
where the LHS gives the position of the center of mass when the
particles are located at positions $-1,\,z,\,+1$.  Although (7) is a
quintic equation for $z$ it has only one solution in the physically
allowed interval $-1<z<1$ and thus $\tau_h$ is unique.  In particular,
if $m_1=m_2$ we have $\tau_h=1$.

\subsection{The Poincar\'e section}
For a comprehensive study of system (1) we must define the initial
values (which define the Poincar\'e surface of section) so that in
principle {\it all} orbits can be included.  For a compact system it
is easy to choose a suitable section, but when open trajectories are
possible the situation can be more involved. Note also that since our
system is open many trajectories will hit the Poincar\'e section only
a few times, while in a closed system this will happen repeatedly.

The location of the section is defined by the particular value of the
ratio $\tau$ defined in (5) above. As coordinates on this section we
introduce the mean value $R$ of the distances $q_1,\;q_2$, and an
angle $\theta$ specifying the distribution of kinetic energy between
the particles. We will now discuss this in detail.

The value of distance ratio $\tau$ that defines the surface of section
must be chosen in a particular way in order to ensure that every
trajectory crosses the surface.  Due to the attractive nature of the
interaction every trajectory (with the exception of the above
mentioned homographic solution) will have pairwise collisions between
particles 1 and 0 and also between particles 0 and 2 (in fact for most
orbits there will be several such points). Thus at a certain time we
will have $q_1=0$ and another time $q_2=0$, and between these times
the ratio (5) will assume all values between 0 and $\infty$. Only for
the homographic solution this ratio will have a fixed value $\tau_h$.
Thus to make sure all trajectories pierce the Poincar\'e section at
least once we will define the location of the Poincar\'e section by
the condition $\tau=\tau_h$.

As the first coordinate on the section we take half the distance
between the outermost particles
\begin{equation}
R={\textstyle\frac12}(q_1+q_2)|_{\tau=\tau_h}.
\end{equation}
When $\tau=\tau_h$ the energy is $E=T-\frac{1}{R}\left[\frac{m_0m_1}{1+z}+
\frac{m_0m_2}{1-z}+\frac{m_1m_2}{2}\right]$, and for a fixed negative
energy the maximum value of $R$ is obtained from this expression  when
the kinetic energy $T$ vanishes:
\begin{equation}
R_{max}=\frac1{|E|}\left[\frac{m_0m_1}{1+z}+
\frac{m_0m_2}{1-z}+\frac{m_1m_2}{2}\right].
\end{equation}

For the remaining coordinate on the Poincar\'e section we must choose
something related to the momenta, but we do not want the kinetic
energy depend on this quantity. Thus we diagonalize the kinetic part
$T$ of $H$ in (7) by writing it as a sum of squares
$T=(ap_1+bp_2)^2+(cp_1+dp_2)^2$.  We can then parametrize $p_i$ in the
required way as
\begin{equation}
ap_1+bp_2=\sqrt{T}\cos(\theta),\quad
cp_1+dp_2=\sqrt{T}\sin(\theta),
\end{equation}
where $\theta$ is the remaining coordinate of the Poincar\'e section.
The constants $a,b,c,d$ are not defined uniquely, but there is a one
parameter family of ways to write T as a sum of squares.  To fix this
final free parameter we impose one more constraint: the above
mentioned homographic solution should be located on the line
$\theta=0$. This fixes the diagonization.

In practice the computation of the constants $a,b,c,d$  proceeds as
follows: If we use $\dot q_i$ instead of momenta, we may write
\begin{equation}
a_{11}\dot q_1+a_{12}\dot q_2=\sqrt{T}\sin(\theta),\quad
a_{21}\dot q_1+a_{22}\dot q_2=\sqrt{T}\cos(\theta)
\end{equation}
and choose the constants $a_{ij}$ such that for $\theta=0$ we
have $\dot q_1/\dot q_2=q_1/q_2=\tau_h$.
When we also require that the kinetic
energy is independent of $\theta$,
a straightforward calculation gives the following set of
formulae from which $a_{ij}$ can be determined:
\begin{eqnarray}
      r_1~&=&1+z,\quad r_2~=1-z \nonumber \\
      A~&=&{\textstyle\frac12}\frac{m_1(m_0+m_2)}{m_1+m_0+m_2},\quad
      B~={\textstyle\frac12}\frac{m_2(m_0+m_1)}{m_1+m_0+m_2},\quad
      C~=\frac{m_1m_2}{m_1+m_0+m_2} \nonumber \\
  c_1~&=&{\textstyle\frac14}(4AB-C^2)/(Ar_1^2+Br_2^2+Cr_1r_2) \nonumber \\
      a_{11}&=&+r_2\sqrt{c_1},\quad a_{12}=-r_1\sqrt{c_1},\quad
      a_{21}=\sqrt{A-c_1r_2^2},\quad a_{22}=\sqrt{B-c_1r_1^2}.
\end{eqnarray}
Since the derivatives $\dot q_i$ and the momenta $p_i$ are related by
\begin{equation}
p_1=2A\dot q_1+C \dot q_2, \quad
p_2=2B\dot q_2+C \dot q_1,
\end{equation}
the constants $a,b,c,d$ are finally determined by comparing (11) and (10).

If $m_1=m_2(=:m_u)$ the formulae simplify considerably and we have
\begin{eqnarray}
\tau_h=1,\quad & &R=q_1|_{\tau=1}, \nonumber \\
{\textstyle\frac12}\sqrt{{\textstyle\frac1{m_u}}+
{\textstyle\frac2{m_0}}}\,(p_1-p_2)= \sqrt{T}\sin(\theta),& &\quad
{\textstyle\frac12}\sqrt{{\textstyle\frac1{m_u}}}\,(p_1+p_2)=
\sqrt{T}\cos(\theta),
\end{eqnarray}
and if $E<0$ we have also $0<R<\frac{5}{2|E|}$.  For this mass
relation it is sufficient to take $0<\theta<\pi$, since
$\theta\to2\pi-\theta$ implies $1\leftrightarrow2$, which is a
symmetry on the Poincar\'e section.  Furthermore, if we change $\theta
\to \pi-\theta$ we get the same trajectory with momenta reversed and
indices changed $1\leftrightarrow 2$.  Since $q_1=q_2$ on the section
these $\theta$-values do in fact correspond to the future and past of
the {\it same} trajectory.  For other mass values these symmetries do
not exist and we must compute twice as many trajectories.

In summary, by using the scaling invariances of the Hamiltonian we can
fix the energy and the sum of masses, which leaves two essential
constants in the Hamiltonian. By the canonical change of coordinates
(2,3) we can eliminate the center of mass motion and reduce the
system to a two dimensional one (4).
The Poincar\'e section is then defined by the values of
$R$ and $\theta$ at $\tau=\tau_h$ as defined in (5-8,10-14).

\subsection{The numerical method}
The Hamiltonian (4) is regularized by the Aarseth-Zare [16] method.
The one-dimensional form of this method consists simply of the point
transformation $q_i=Q_i^2$, which gives the new canonical momenta
$P_i=2Q_i p_i $. Substituting this together with the time
transformation $dt/ds=q_1q_2$, where $s$ is a new independent variable
(note that $dt/ds\ge0$), and applying Poincar\'e's transformation we
have the regularized Hamiltonian $\Gamma = q_1q_2(H-E)$ in the form
\begin{eqnarray}
 \Gamma &=&
{\textstyle\frac18}\left\{\left(\frac1{m_1}+\frac1{m_0}\right)P_1^2Q_2^2+
\left(\frac1{m_2}+\frac1{m_0}\right)P_2^2Q_1^2-
{2\over{m_0}}P_1P_2Q_1Q_2\right\} \nonumber \\
& &\quad -m_0m_2Q_1^2 -m_0m_1Q_2^2
-m_1m_2\frac{Q_1^2Q_2^2}{Q_1^2+Q_2^2}-Q_1^2Q_2^2E.
\end{eqnarray}
Here $E$ is the constant numerical value of the energy, calculated
from initial values.

The equations of motion derived from (15) are $ Q_i'= { {\partial
\Gamma}/ {\partial P_i} }$, $ P_i'=-{ {\partial \Gamma}/ {\partial
Q_i} } $, $t'=Q_1^2Q_2^2$, where differentiation with respect to the
new independent variable $s$ has been denoted by a prime and an
equation for the time has been added. With these equations the
numerical integration of pairwise collisions is no more difficult than
that of the harmonic oscillator (into which equations of motion of the
colliding pair reduce asymptotically).

The initial values in our calculations are chosen on the surface of
section discussed in Sec. 2.2.  When we integrated numerically the
evolution of a particular orbit we checked for the changes of the sign
of the quantity $q_1-\tau_h q_2$ and used it to determine accurately
the point on the surface of section.

The calculations were done using the Bulirsch-Stoer method [19], which
uses rational function extrapolation to zero steplength from results
obtained by the midpoint rule with several different (sub-)steps.  The
method estimates its error from the various extrapolation outcomes and
adjusts the stepsize by comparing the error estimates with a specified
tolerance for the one-step error.  The tolerance we used for relative
precision was $10^{-12}$.  The value of the quantity $\delta E/L$,
where $\delta E$ is the error in energy and $L$ is the Lagrangian (sum
of the absolute values of kinetic and potential energy), was less than
$10^{-12}$ for two thirds of the computed orbits, while the maximal
values were between $10^{-9}$ and $10^{-10}$.  The fraction of cases
in this worst interval was less than one out of 2000. Thus we conclude
that a high precision was ensured by the used error tolerance.

\section{Types of motion}
In this section we first discuss the types of motion at a general
level and then survey a selection of typical three-body orbits.

\subsection{General taxonomy}
For a first rough classification of orbit types of the rectilinear
three-body systems one can distinguish between those that stay bound
and those that break up.  If the total energy is negative the breakup
result is always a bound binary and a single unbound particle. A
connection between the past and future was provided by Hopf [20], who
showed that the system actually stays bound forever or breaks up in
both directions of time, with the exception of a set of measure zero.
(Since the exceptional set has measure zero it cannot appear in a
numerical study, even though it is not empty [21].) In a system with
positive total energy also a total breakup into three unbounded
particles is possible (but not necessary).  In a more detailed
classification one must separate the `bound forever' set into periodic
and quasi-periodic orbits, and the `breakup' set into fast scattering
(often called `non resonant' scattering [4]) and long interplay
(`resonant interactions').

Among the bounded orbits there are periodic orbits, of which a special
one is known as the {\it Schubart orbit }.  This orbit is the periodic
orbit which has the shortest period and it is thus the simplest
periodic orbit in the system under consideration.  It was found
numerically by Schubart [10] in 1956 for the case of equal masses.
Later the corresponding orbits for other masses were studied by
H\'enon [11] and by us (Paper III).  If the Schubart orbit is linearly
stable (stability is discussed further in Sec. 5) then it is usually
(in the absence of destructive resonances) surrounded by a finite
region of quasi-periodic orbits.  This follows from the KAM-theory.

Since the system is two dimensional the quasiperiodic motion has two
basic frequencies. When the ratio of these frequencies is a rational
number the motions resonate. We will later see that
especially the 1:2 and  1:3 resonances have strong influence on the
dynamics.

If the orbit is not bounded for all times the triplet will break up in
the past and in the future.  The initial configuration of such an
orbit can be characterized by the the label of the free particle and
by the binding energy of the other two particles, similarly for the
final configuration.  We can have either an exchange interaction
(where the free particle is different) or backscattering (where the
free particle is the same one in both directions of time).
Furthermore, in the latter case the free particle may be the lighter
or the heavier one of the outside particles.  Interesting features
will be shown to be related to this classification.

In addition to characterizing the initial and final state constituents
a scattering system can be characterized by the `dwell time' (also
called `interplay time'), which is defined as the time during which
all the three particles stay `close together'. To make this vague term
more precise we need a working definition for the moments of breakup
in the future and in the past and then define dwell time as their
difference. In Paper I we defined the moment of breakup as the formal
pericenter time for the asymptotic two-body orbit of the escaping
(entering) particle with respect to the center of mass of the
remaining binary.  Another possible definition would be the time of
last (first) crossing of the Poincar\'e section. In practice these
definitions work much the same way and their numerical values are
close to each other.

As will be shown later, the scattering orbits can be classified into fast
scattering and chaotic scattering. For fast scattering the dwell-time
defined by difference of pericenter times is nearly zero and sometimes
even negative, while the difference of last and first crossing of
Poincar\'e section is always a small positive number. We will often refer to
the fast scattering orbits as `orbits of zero interaction time'.
For fast scattering the dwell time and the initial and final states
depend smoothly on the initial point.
For chaotic orbits the dwell time can be arbitrarily large, and it as
well as the final state depend sensitively on the starting point on
the Poincar\'e section.

One can also arrive at essentially the same classification to fast and
chaotic scattering by counting the number of Poincare map points
produced by the orbit. Each full orbit has a first and a last
Poincar\'e map point, which may be called the {\it entry} point and
the {\it exit} point. As was shown in Paper I, these entry and exit
points are located in clearly defined regions of the Poincar\'e
section. For each mass configuration there is a minimum number of
crossings, which is larger for smaller center mass. For chaotic orbits
the number of sections is characteristically much larger than the
minimum and it depends sensitively on the starting point.

We will show later that the different types of orbits, fast
scattering, chaotic scattering, and quasi-periodic, are located in
separate regions in phase-space (and on the Poincare map).

\subsection{Typical trajectories}
To get the first impression of the different orbit types
we look now in more detail at a selection of individual orbits given
in Fig.\ \ref{fig3}. The orbits are arranged in increasing dwell time.
With the exception of the quasiperiodic orbits g) and h) these
illustrations are complete in the sense that at both ends the final
states are shown and the parts will just fly apart without any further
threebody interaction.

In Fig.\ \ref{fig3}a we have an orbit with three isolated particles at
both $t=-\infty$ and $t=+\infty$. Here the energy is positive (as it
always is for systems which break up into three unbounded particles)
and the individual trajectories are almost straight lines outside the
interaction region.  During interaction there is some energy transfer.
The trajectories seem to be translated back somewhat, indicating a
small negative dwell time, which is natural for an attractive
interaction.

Figs.\ \ref{fig3} b and c illustrate fast scattering with a binary and
a single particle at both $t=-\infty$ and $t=+\infty$. This is the
typical situation for fast scattering at negative energy but is also
possible for positive energy. In Fig.\ \ref{fig3}b the single particle
is different at $-\infty$ and $+\infty$, in Fig.\ \ref{fig3}c it is
the same. In both cases some energy transfer takes place, as can be
seen from the changing oscillation time of the binary. [For positive
energy it is also possible to have a binary + single particle in one
direction of time and three separated particles in the other, c.f.
paper II for further illustrations.]

Figs.\ \ref{fig3} d-f show typical chaotic orbits for various mass
choices.  The sensitivity to initial values in Fig.\ \ref{fig3}d is
particularly illuminating: When the single particle makes a long
detour and thereafter again interacts with the binary the outcome of
that interaction depends on the relative phase of the binary
oscillation. The phase in turn depends sensitively on how long the
detour took, i.e. how fast the binary and the single particle started
to separate. In Fig.\ \ref{fig3}e the three body system breaks up
after an almost three-body collision resulting with a very tight
binary plus an equally energetic single particle. This outcome would
change drastically with even a small change in the starting
configuration.

Finally in Figs.\ \ref{fig3} g and h we have typical quasiperiodic
orbits. In Fig.\ \ref{fig3}g the masses are equal. The orbits seems to
be near a 1:2 resonance, at which the Schubart orbit becomes unstable
(see Sec. 5).  The configuration in Fig.\ \ref{fig3}h seems to be
close to a 1:5 resonance.

\section{Dwell time charts}
In order to get a comprehensive overall picture of the possible
motions it is not sufficient to look at particular orbits.  In Papers
I and II, where we studied the equal mass case in detail, we computed
the orbits for a rather dense lattice of starting points on the
$(R,\theta)$-plane.  Here we use a $100\times360$ linear lattice for
the initial values on the Poincar\'e section
\begin{equation}
 R_\nu =(\nu-{\textstyle\frac12})\times 0.025, \quad \nu=1,\dots, 100; \quad
\theta_\mu=(\mu-{\textstyle\frac12})\times 1^o, \quad \mu=1,\dots, 360.
\end{equation}
[If the outside masses are equal (papers I and II) there is a
reflection symmetry (see Sec. 2.3) and only the range
$\mu=1,\dots,180$ needs to be studied.]  We calculated the orbits
starting at these points $(R_\nu,\theta_\mu)$ until the final type of
motion was evident.

Now with 36000 orbits calculated one must worry about data
presentation. In I and II we looked at how the trajectory type and the
dwell time depended on the initial values.  The data was presented by
drawing a box around the initial point on the Poincar\'e section and
coloring it with a shade of gray according to how the orbit behaved.
Here we use the same method and first recall the dwell time data for
the equal mass case of Paper I and for comparison present similar new
data for a nearby mass configuration.

Fig.\ \ref{fig4}a shows the dwell time for each of the initial values
(16) [Energy has been scaled to $-1$].  The darker the small square is
the longer it takes for the orbit starting at the center of the square
to break up (since energy is negative the break up result can only be
a single particle + a binary). The figure is symmetric across $180^o$,
because $m_1=m_2$.  The `one directional' dwell time shown in
Fig.\ \ref{fig4}a means the time starting from an initial value (16)
until the system breaks up. The two directional dwell time of Fig.\
\ref{fig4}b is the sum of the two dwell times calculated from the same
starting point into both directions of time, i.e. this is a total time
for the triple interplay for the orbit that at some time passes
through the center of the square. This figure is symmetric across
$90^o$.

The most prominent feature of Figs.\ \ref{fig4} is the way the Poincar\'e
section is divided into three well defined regions:

1) Around $R=0.8,\,\theta=90^o,\,270^o$ there is a black region where
the orbits are quasiperiodic and the three-body system stays together
forever.  We call this the {\it Schubart region}, because at its
center there is the periodic Schubart orbit.  The discretization does
not reveal the nature of the region's boundary, i.e. whether it is
fractal of smooth. We have studied this in more detail but did not
find any signs of fractality.  [The situation is probably different
for other mass values.]

2) The Schubart region is surrounded by a grayish area that extends to
the boundaries of the section.  This region seems to have no apparent
structure, i.e. there no correlation between the dwell times of
neighboring initial values. We call this region chaotic, because the
trajectories depend sensitively on the initial values, which is one
signature of chaos. The origin of this chaotic behavior was discussed
in connection with Fig.\ \ref{fig3}d.

The statistical distribution of dwell times (paper I, Fig. 9) shows
an excess of long time orbits when compared to the usually obtained
exponential decay [1]. This is due to the long range interaction,
which makes arbitrarily long detours quite common; such detours are
completely absent if the interaction has finite range.

3) The third region consists of the white `scallops' at the lower part
of the figure.  When the initial values are chosen from this region
the system breaks up with at most 2 further collisions. The leftmost
scallop turns out to be the exit region, i.e. if the initial values
are chosen from it the system breaks up with no further crossing of
the Poincar\'e section.  [The shape of the exit region can be
approximated analytically [22].]

Even for a chaotic scattering orbit it is necessary that its last
point in the Poincar\'e section is also in the leftmost scallop, which
is the exit region discussed above. [This is because the exit point
has zero dwell time in the future and long dwell time in the past.]
Indeed, in the folded Fig.\ \ref{fig4}b there is a gray spike inside the
leftmost scallop.  This is where the chaotic orbits pierce the
Poincar\'e section the last time before they break up, the rest is for
orbits whose total interaction time is practically zero.

In Fig.\ \ref{fig5} we have charted the dwell time in the same way for
the mass choice $(0.9,\,1,\,1.1)$. The forward orbit chart in Fig.\
\ref{fig5}a is no more symmetric and neither is the full orbit chart
Fig.\ \ref{fig5}b.  The masses are almost equal but the mass
configuration is nevertheless such that the Schubart orbit is unstable
(see next section).  Indeed, there is no black Schubart region left.
Its remnants show up as darker gray, indicating that the orbits there
do break up, but not so fast.  These changes look rather drastic given
that the masses do not differ much from the equal mass case.  In Fig.\
\ref{fig5}a the chaotic region is still for the most part without
structure, but there also seem to be some continuous lighter regions.
The fast scattering scallops are still there, but clearly modified
from the case of equal masses.

In papers I and II we used the same kind of charting technique to show
the type of the initial and final state (Fig.\ \ref{fig2}b in paper I,
and Figs. 1-3 in paper II) and their binding energy (Fig. 3 in I). For
negative energy all but the quasiperiodic orbits have initial and
final states consisting of a single particle and a binary.  The orbits
were then classified according to whether it was particle 1 or 2 that
was the single particle. In the negative energy region the figures
show a structure similar to Fig.\ \ref{fig4}: In the chaotic region
the final state depends sensitively on the initial conditions, while
within each scalloped fast scattering region it does not change.

Let us also discuss here how the transition to chaos takes place.
First consider the case when the initial point moves from the fast
scattering region to the chaotic scattering region. As we move inside
a scallop towards its border the asymptotic speed of the escaping
particle changes smoothly and approaches zero when the initial point
approaches the boundary. Thus at the boundary we have an
(asymptotically) parabolic escape. Just at the other side of the
boundary the escape is replaced by an `ejection without escape', i.e.\
the nearly escaping particle completes a very long elliptic orbit
before returning to strong interaction with the other particles. While
this ejected particle is making its journey the binary completes a
large number of periods and the nature of the subsequent threebody
interaction depends sensitively on the escape velocity. This implies
in fact that on the chaotic side of the line of parabolic escape there
is a clustering of singular triple collisions.

The transition from quasiperiodic bounded system to chaotic scattering
is different in nature. If we start from a quasi-periodic motion and
move the initial point towards the boundary of the Schubart region we
arrive at a `broken' KAM-torus and eventually the particle finds its
way out of the quasi-periodic behavior.  Near the border some
quasi-periodic looking behavior is still observable. For some mass
values the Poincar\'e sections discussed later show unmistakably the
cross sections of broken tori around the Schubart orbit.

\section{The mass triangle and the stability of the Schubart orbit}
In the following sections we will look more closely on how the
previously observed characteristics change when the masses are
arbitrary.  We will see that the three basic regions identified above
will for the most part remain, although in a distorted form, but for
some mass values the Schubart region vanishes completely.  This is
determined by the stability of the Schubart orbit, which we will
discuss next.

Because of scaling invariance we can normalize the masses so that
\begin{equation}
m_1+m_0+m_2=3,
\end{equation}
and then parametrize the mass plane by $a$ and $b$ as follows:
\begin{equation}
m_1=1-a-b,\, m_0=1+2a,\, m_2=1-a+b,\text{ where }
-{\textstyle\frac12}\le a\le 1,\, a-1\le b \le 1-a.
\end{equation}
The mass triangle is given in Fig.\ \ref{fig6}. The equal mass case
discussed in Papers I and II is marked with a cross. For computations
(Paper III) we discretized the mass triangle by
\begin{equation}
a=0.01\cdot I_a,\quad b=0.01\cdot I_b,\quad I_i \text{ integers}.
\end{equation}
For compatibility with our earlier published work we use this same
notation here, although we mostly discuss the case were the mass
indices $I_i$ are multiples of ten.

One of the most prominent features in the equal mass case was the
Schubart region. As the masses change we would expect this region to
change and perhaps even vanish. [A preview of this is provided by
comparing Figs.\ \ref{fig4} and \ref{fig5}.] At the center of the
Schubart region is the periodic Schubart orbit, which presumably would
be the last orbit to destabilize.

In Paper III we analyzed the Schubart orbits by finding numerically
the periodic solution (it exists for all mass values) and calculating
its stability. The stability is related to the eigenvalues of the
transition matrix $\bbox{A}$ of the Poincar\'e map; $\bbox{A}$ gives
the difference in the next Poincar\'e maps point as a function of the
difference in the initial point: $d\bbox{Y}=\bbox{A}d\bbox{Y}_0$,
where $\bbox{Y}=(R,\theta)$.  Since the system is Hamiltonian the
matrix $\bbox{A}$ is symplectic and the product of its eigenvalues is
$=1$. If the eigenvalues are real, one of them has magnitude $>1$ and
the orbit is unstable. If the eigenvalues are complex conjugates, they
are both on the unit circle and the Schubart orbit is linearly stable.
Furthermore, if the eigenvalues have the form $\lambda=\exp(i\theta)$,
with $\theta=2\pi/m$, $m$ an integer, then we have a $1:m$ resonance
in the two periodicities associated with the motion (in the immediate
neighborhood of the Schubart orbit).

In Fig.\ \ref{fig6} we have given as gray those regions for which the
Schubart orbit is stable under perturbations that nevertheless keep
the system rectilinear (longitudinal perturbation) [11, III].  (For a
discussion of perturbations that allow the masses to move on a plane
(transverse perturbations), see [11, III].)  It is interesting to note
that the equal mass case is quite close to the fairly large unstable
region.

In Fig.\ \ref{fig6} we have also given one eigenvalue of the map $\bf
A$ at selected points using a dial. As expected the region of
instability is bounded by the 1:2 resonance where both eigenvalues
reach the point $-1$ on the real line (`reading of the dial' is
$180^0$).  Thus the destabilization is a typical one: the eigenvalues
move from the unit circle through the point $-1$ to the real axis. The
figure shows nicely how in the stable region the eigenvalues turn
smoothly as the masses change.

We will later show that the mass regions where the Schubart orbit has
longitudinal stability correlate also with a certain characteristic of
the final states of chaotic orbits.

\section{Poincar\'e sections for various masses}
To get a more detailed look at the changes due to mass variation we
have computed the Poincar\'e sections for 105 mass values
\begin{equation}
I_a=-40,-30,\dots,90,\quad I_b=0,10,\dots,90-I_a.
\end{equation}
In each case we used $180$ initial values
\begin{equation}
\theta=5,15,\dots,175,\quad R/R_{max}=0.05,0.15,\dots,0.95,
\end{equation}
and integrated the trajectories in both directions of time (this is
equivalent to computing trajectories forward only, but allowing
$\theta$ to range from $0^o$ to $360^o$).  [Note that for general mass
values the many useful symmetries present in the equal mass case no
longer exist.]

The 105 Poincar\'e sections are arranged in the form of the mass
triangle as shown in Fig.\ \ref{fig7}. This shows the contents of the
more detailed Figs.\ \ref{fig8} a-g, but it does also give an overall
picture of the basic features discussed in the previous two sections.

1) Let us start from the Schubart region. In general its size
increases as the central mass decreases (going down in the mass
triangle) and it gets tilted when the ratio of the outside masses
change. In the mass region where the Schubart orbit is longitudinally
unstable (c.f. Fig.\ \ref{fig6}) the whole Schubart region has
vanished, as expected.  There are also other mass values with
vanishing Schubart regions.  Some of them are at the 1:3 resonance
which is located near $I_a=40$ line illustrated in Figs.\ \ref{fig8}
e,f (the position of the 1:3 resonance depends weakly on $b$ as well).
There is also another curve in the linear stability region where the
Schubart region vanishes, it starts from the lower right hand corner
and passes close to points $(I_a,I_b) =(-40,110),\, (-30,60)$ [Figs.\
\ref{fig8} d,c, respectively.] This corresponds also to a 1:3 resonance.
However, this latter curve of vanishing Schubart regions does not seem
to extend all the way to $I_b=0$.  Fig.\ \ref{fig9} shows a magnified
Poincar\'e section for the mass parameters $(40,0)$. There is a tiny
Schubart region, and around it trajectories that diverge with period
three.

2) The Schubart orbit is surrounded by the chaotic scattering region,
even when the Schubart region vanishes (either by linear instability
or the 1:3 resonance). The boundary between the chaotic scattering
region and Schubart region (when the latter exists) is not always so
clear as it was in the equal mass case. As is expected from general
theory, the tori surrounding the Schubart orbit do sometimes break up
and form islands in the sea of chaotic scattering.  Rather large
islands appear e.g. in $(-20,40)$ [Fig.\ \ref{fig8}b].

Let us look more closely how these islands appear as the mass
parameter $I_b$ changes in the $I_a=-40$ region [bottom of Fig.\
\ref{fig8} b,c].  A detailed illustration of this is provided in
Fig.\ \ref{fig10}.  In Fig.\ \ref{fig10}a ($I_b=59$) the 9 islands are
well inside a clear torus, then the islands seem to move away from the
center and finally in Fig.\ \ref{fig10}c ($I_b=53$) the islands are
inside the chaotic sea.  The same process takes place when we approach
the above mentioned $(-20,40)$ from $(-20,30)$ or $(-30,40)$ [Fig.\
\ref{fig8}b].

The dark crescent in $(20,60)$ [Fig.\ \ref{fig8}g] is intriguing.
It is located in the region of unstable Schubart orbit, but
nevertheless there are some very long-lived orbits, whose fate we
could not determine. It seems that the angle of homoclinic
intersection is here close to zero. This region needs further
investigation.

3) Finally, let us look at the scallops of fast scattering.  The
overall behavior is clear from Fig.\ \ref{fig7}: The scallops at the
bottom of each Poincar\'e section get smaller and more numerous as the
central mass decreases going down in Fig.\ \ref{fig7}, and when we go
towards right (towards more lopsided mass configurations) the scallops
increase in size and get more asymmetric.

The number of scallops indicates the minimum number of collisions
needed in the fast region.  This can be seen as follows: For
$\theta$-values near $\pi$ the masses 1 and 2 are moving towards mass
0 ($\dot q_1 \sim\dot q_2 <0$) and thus the rightmost scallop is the
entry region.  With each collision the speed of approach gets smaller
and the corresponding points in the Poincar\'e section move left.
Eventually the masses 1 and 2 start to recede and the exit takes place
from the leftmost scallop where $\theta$ is near 0 ($\dot q_1
\sim\dot q_2 >0$).  The number of scallops increases as the center
mass decreases, because when we have a smaller intermediary mass it
must make more collisions with the other two in order to transfer
enough momentum to make the approaching masses recede.

Between the scallops the spikes of chaotic scattering region reach the
$R=0$ line, as was observed before.  What was not visible in the
dwell-time charts is that the spikes seem to continue inside the
chaotic region as slightly darker lines. These darker lines seem to
reach from the $R=0$ line the all the way to the corners of the
Schubart region.  This is clear e.g. for $I$ values
$(-20,40),\,(-20,50)$ [Fig\ \ref{fig8}b], but can also be seen
elsewhere. This suggest that the stability of the Schubart orbit has
global influence, another manifestation of this is discussed next.

\section{Distribution of the trajectory types}
  From the previous it seems that the Schubart region is a dominating
feature of the system. At the end of the previous section we noted
that there are faint lines that extend from the corners of the
Schubart region all the way to the $R=0$ line. In this section we
present another puzzling observation about the global influence of the
stability of the Schubart orbit.

In the Figs.\ \ref{fig11} a-e we have plotted the initial and final
states types using the same method as in Figs.\ \ref{fig4} and
\ref{fig5}. That is, at the starting point of an orbit we have drawn a
mark according to the fate of the orbit in the future and in the past.
We have used only a $10\times 18$ initial value grid for each mass
configuration so that we could combine the individual charts as in
Fig.\ \ref{fig7}. For each property we have constructed a different
overview chart. In each figure we have also drawn borders around the
region where the Schubart orbit is unstable under longitudinal
perturbation, see Fig.\ \ref{fig6}. [In a very few cases the system
was still evolving when the computations were stopped, these cases are
omitted.]

Fig.\ \ref{fig11}a we have marked the quasiperiodic orbits with a
small triangle while all other orbits were left blank.  The region
where the Schubart orbit is unstable is of course empty, except in the
subchart $(I_a,\;I_b)=(20,60)$, where an orbit survived as a bound
system for more than 40000 Poincar\'e section crossings (at which
point the computation was halted). This orbit is from the crescent
shown in Fig.\ \ref{fig8}g.  This particular mass triplet is very
close to the border of the Schubart orbit instability region, but a
careful analysis shows that it is inside.  The row at $I_a=40$, which
is near 1:3 resonance, is also clearly visible as an empty region
showing that the resonance has destroyed the KAM-tori entirely.  On
some subcharts there are also isolated points which corresponding to
the separate islands; they are more clearly visible in the Fig.\
\ref{fig8} (e.g. Fig.\ \ref{fig8}b $(-20,40)$).

Fig.\ \ref{fig11}b shows the fast scattering orbits, all others were left
blank. There seems to be no particularly interesting structure here.
[For this figure we used a working definition by which the scattering
was termed fast if it had at most 4 points on the Poincar\'e section.
This is not a good definition since the minimum number depends on the
center mass, as was discussed before, and causes the apparent absence
of fast scattering at the bottom of the figure where the center mass
is small.]  In Fig.\ \ref{fig11}c we have plotted those chaotic scattering
orbits for which the single particle is light in both past and future,
there is no interesting structure here either.  This interaction type
is obviously the most common one.  At the leftmost column the outside
particles have equal masses and we have used particle 1 as the lighter
one, as it is elsewhere.

However, when we plot the initial conditions leading to a heavier
single particle, either in the past or in the future (Fig.\
\ref{fig11}d) or both (Fig.\ \ref{fig11}e) some interesting structure
emerges: We see that such scattering orbits seem to be curiously
absent when the Schubart orbit is unstable.  Thus the stability of the
Schubart orbit seems to radiate its effect throughout the chaotic
region, correlating strongly with the ability of the heavier particle
to escape.  We observed the same phenomenon also when we were
calculating the results illustrated in Fig.\ \ref{fig5}. In this much
denser grid the rule also worked without exception: In Fig.\
\ref{fig5} the mass configuration was such that the Schubart orbit was
unstable and for the chaotic scattering orbits we found indeed that
the escaper was always the lighter body.

  From this numerically observed rule we can also get the following
predictive forms (in each case we must of course assume that the mass
configuration is such that the Schubart orbit is unstable and the
total energy of the system is negative):
1) When a binary and a single particle collide the scattering will
take place in minimum time if the incoming
single particle is the heavier one of the outer particles.
2) If the scattering process looks chaotic the lighter one of the
outside particles will eventually be ejected.

This numerically observed correlation with the stability of the
Schubart orbit and the initial and final state type is still waiting
for an analytical explanation.

\section{ Conclusions}
In this paper we have presented data characterizing the behavior of
the rectilinear gravitational three-body motion. This, possibly the
simplest nonintegrable three-body case, nevertheless shows a rich
variety of phenomena which is not yet thoroughly understood.

Scattering systems have been studied actively in recent years, but the
usual three-disk case is quite different from the system studied here
and as a consequence shows different behavior. This is not unexpected,
since when the present system is looked as a 2-dimensional system it
is described by a particle moving in the field of three
attracting walls (Fig.\ \ref{fig2}), furthermore the interaction is
long-range.

Perhaps the most characteristic feature of the present system is the
division of the Poincar\'e section into three regions: The
quasiperiodic Schubart region where the system stays bounded for all
times, the chaotic scattering region with finite nonzero dwell times
that depend sensitively on the initial values, and the `scalloped'
fast scattering region. The three regions can be seen clearly in the
dwell time figures, in the Poincar\'e sections, and in the final state
distributions.

An interesting finding is the way the Schubart region influences the
global picture. In Figs.\ \ref{fig8} we saw the darker lines extending
from the bottom of the Poincare section to the corners of the Schubart
region. In Fig.\ \ref{fig11} we saw how certain types of scattering
processes are completely absent if the periodic Schubart orbit is
unstable.  It is of great interest to find an analytical explanation
for these numerical observations.

\acknowledgements
 One of us (J.H.) would like to thank Roberto Camassa, David K.
Campbell, Greg Forest and Darryl Holm for discussions, and David K.
Campbell for hospitality at the Center for Nonlinear Studies, Los
Alamos National Laboratory, where this work was partially done,
supported by contract W-4705-ENG-36. We would also like to thank the
referees for useful comments.

\begin{figure}
\caption{ The mass configuration. The three masses,
labeled $i=1,\, 0,\, 2$ from left to right (masses $m_i$ at
positions $x_i$) attract each other with the Newtonian force
 $m_im_j{\vert x_i-x_j\vert} ^{-2}$.}
\label{fig1}
\end{figure}

\begin{figure}
\caption{ A typical motion in the $(q_1, q_2)$-plane.
The motion is restricted to the first quadrant.  In this
representation the coordinate axes and the line $q_1+q_2=0$ are
attracting walls. The line $q_1=q_2$ corresponds to the Poincar\'e
section. The plotted trajectory represents a fast exchange scattering
process with equal masses. The arrows show the direction of motion
(reverse motion is also possible).}
\label{fig2}
\end{figure}

\begin{figure}
\caption{ Typical examples of the basic orbit types.
The coordinates of the particles (vertical axis) are plotted as
functions of time (horizontal axis).  The lower curve corresponds to
the motion of the particle number 1 and the upper curve to particle 2.
The orbit of the middle body has been plotted using dashed curve.  a)
A high energy orbit.  [$E=10,\; R=1,\; \theta=10^o$, $m=(0.9, 1,
1.1)$].  b) A fast scattering orbit. This is an exchange interaction
in which the free particle is different at $t\rightarrow -\infty$ and
$t\rightarrow \infty$. [$E=0,\;R=1,\; \theta=45^o$, $m=(0.9, 1, 1.1)$].
c) A fast back-scattering orbit. The free particle is the same at both
directions of time. [$E=0,\; R=1,\; \theta=10^o$, $m=(0.9, 1, 1.1)$].
d) A chaotic orbit with almost equal masses. [$E=-1,\;R=1, \theta=10^o$,
$m=(0.9, 1, 1.1)$].
e) A chaotic orbit with a large central mass. [$E=-1.1766,\; R=1,\;
\theta=90^o$, $m=(0.2, 2.4 ,0.4)$].
f) A chaotic orbit with a lopsided mass choice. [$E=-1.7456,\; R=1,\;
\theta=90^o$, $m=(0.4, 1, 1.6)$].
g) A quasi-periodic orbit in the equal-mass case. [$E=-2.000,\; R=1,\;
\theta=75^o$, $m=(1, 1, 1)$].
h) A quasi-periodic orbit with a small middle mass. [$E=-1.2138,\; R=1,\;
\theta=45^o$, $m=(1.2, 0.2, 1.6)$].}
\label{fig3}
\end{figure}

\begin{figure}
\caption{ Map of the dwell times for the equal
mass system as a function of initial values, a) one directional dwell
times, b) total dwell times for the full orbit.  The shade of gray
corresponds to the length of the triple interaction for the orbit
which started at the center of each box. In both illustration the
integer part of $\ln(1+t)$ is used to determine the shade of gray: we
used white when the integer part of $\ln (1+t)=0$, and black when it
was $\ge 10$.  Note the three different regions: black quasiperiodic
region around the Schubart orbit, the grayish chaotic region, and the
vwhite fast scattering region.}
\label{fig4}
\end{figure}

\begin{figure}
\caption{ Map of the dwell times for the mass choice
$(0.9,\,1,\,1.1)$, as a function of initial values, a) one directional
dwell times. b) two directional dwell times.  The grayness scale is
the same as in the previous figure.  Note that the Schubart region has
disappeared although some traces are still visible as a darker
distorted area.}
\label{fig5}
\end{figure}

\begin{figure}
\caption{ The mass triangle. The mass indices
$I_a,\;I_b$ are, as introduced in the text, 100-fold values of the
mass-triangle coordinates $a,\;b$. In the gray region the Schubart
orbit is linearly stable. A dial is placed at certain values of the
mass indices, it shows one of the eigenvalues of the Poincare\'e map
at the Schubart orbit of that particular mass configuration. If the
eigenvalue has unit magnitude we plot the one with positive imaginary
part with a dial ending with a black dot. This happens in the grey
stability region. When the eigenvalues are on the real line  we plot
the one with the smaller magnitude with a cross. This happens in the
white instability region.  }
\label{fig6}
\end{figure}

\begin{figure}
\caption{ A global overview of the Poincar\'e maps
for various mass choices. This also indicates how the various parts of
the next figure are to be combined.}
\label{fig7}
\end{figure}

\begin{figure}
\caption{ Poincar\'e maps for various mass values.
For each map we used a $10\times18$ lattice of initial values.
The indices (e.g. $(-10,0)$) in the lower left corner of each figure
indicate the mass-indices $(I_a,I_b)$.}
\label{fig8}
\end{figure}

\begin{figure}
\caption{ A detail of the Poincar\'e section for
$m=(0.6,\,1.8,\,0.6)$ near the 1:3 resonance of the Poincar\'e map.
 The initial values were chosen from a
box around the point $\theta=90^o,\,R=0.82$. }
\label{fig9}
\end{figure}

\begin{figure}
\caption{ A detailed look on the behavior of the islands
as the masses change, the masses are $(0.81,\,0.2,\,1.99)$ for a),
$(0.84,\,0.2,\,1.96)$ for b), and $(0.87,\,0.2,\,1.93)$ for c).}
\label{fig10}
\end{figure}

\begin{figure}
\caption{ Characterization of orbits according to the
type of motion.
{a)} Quasi-periodic:
The initial value points in the $10\times 18\;\;(R,\theta)$-grid
that gave rise to quasi-periodic motion. The region where
the Schubart orbit is unstable is empty, as is the row at $I_a=40$ near
the 1:3 resonance.
{b)} Fast:
Here we have marked fast trajectories, which were here defined as having
at most 4 points on the surface of section for the entire
orbit from $-\infty\le t \le +\infty$.
{c)} Light$\rightarrow$Light:
Here we plot the starting points of those orbits which were
classified as chaotic and for which  free particle is the lighter one
of the two possible escapers (=outside particles) in both direction of time.
{d)} Heavy$\rightarrow$Light \& Light-Heavy:
This is the exchange interaction. Observe the complete absence of this
orbit type in the region where the Schubart orbit is linearly
unstable.
{e)} Heavy$\rightarrow$Heavy:
One notes that transitions from heavy free particle back to the same
heavy free particle are in general rather rare and completely
impossible when the Schubart orbit is unstable.}
\label{fig11}
\end{figure}

\end{document}